\documentclass[%
reprint,
superscriptaddress,
 amsmath,
amssymb,
aps,
prl,
numerical,
]
{revtex4-1}

\usepackage[USenglish]{babel} 
\usepackage [table]{xcolor}

\usepackage{graphicx} 

\usepackage{amsmath}
\usepackage{amsthm}
\usepackage{amsfonts}

\usepackage{float}
\usepackage{textcomp}
\usepackage{graphicx}
\usepackage{dcolumn}
\usepackage{bm}
\usepackage{hyperref}
\usepackage[mathlines]{lineno}

\begin{document}


\preprint{APS/123-QED}

\title{Slow noise processes in superconducting resonators}

\author{J. Burnett}
\affiliation{ National Physical Laboratory, Hampton Road, Teddington, TW11 0LW, UK}
\affiliation{ Royal Holloway, University of London, Egham, TW20 0EX, UK}
\author{T. Lindstr\"{o}m}
 \email{tobias.lindstrom@npl.co.uk}
\author{M. Oxborrow}
\affiliation{ National Physical Laboratory, Hampton Road, Teddington, TW11 0LW, UK}
\author{Y. Harada}
\author{Y. Sekine}
\affiliation{NTT Basic Research Laboratories, 3-1 Morinosato Wakamiya, Atsugi, Kanagawa 243-0198, Japan}
\author{P. Meeson}
\affiliation{ Royal Holloway, University of London, Egham, TW20 0EX, UK}
\author{A. Ya. Tzalenchuk}
\affiliation{ National Physical Laboratory, Hampton Road, Teddington, TW11 0LW, UK}

\date{\today}

\begin{abstract}

Slow noise processes, with characteristic timescales $\sim$1s,  have been studied in planar superconducting resonators. A frequency locked loop is employed to track deviations of the resonator centre frequency with high precision and bandwidth.  Comparative measurements are made in varying microwave drive, temperature and between bare resonators and those with an additional dielectric layer. All resonators are found to exhibit flicker frequency noise which increases with decreasing microwave drive. We also show that an increase in temperature results in a saturation of flicker noise in resonators with an additional dielectric layer, while bare resonators stop exhibiting flicker noise instead showing a random frequency walk process.
\end{abstract}

\maketitle




Slow fluctuations in charge sensitive devices have been frequently examined over the past few decades\cite{Koch1980}. Recently, their effects were indirectly observed in superconducting qubits\cite{Grabovskij2012} with supporting theoretical work\cite{Faoro2012} linking them to the presence of two level fluctuators (TLFs)\cite{Shnirman2005}. We present measurements directly probing these slow noise processes in superconducting resonators using a high bandwidth feedback technique with Hz level resolution\cite{Lindstrom2011}. Feedback maintains a lock to the resonator centre frequency indefinitely, providing a direct measure of the nature of slow fluctuations and their behavior in varying temperature, microwave drive and TLF density.
\begin{figure*}
\includegraphics{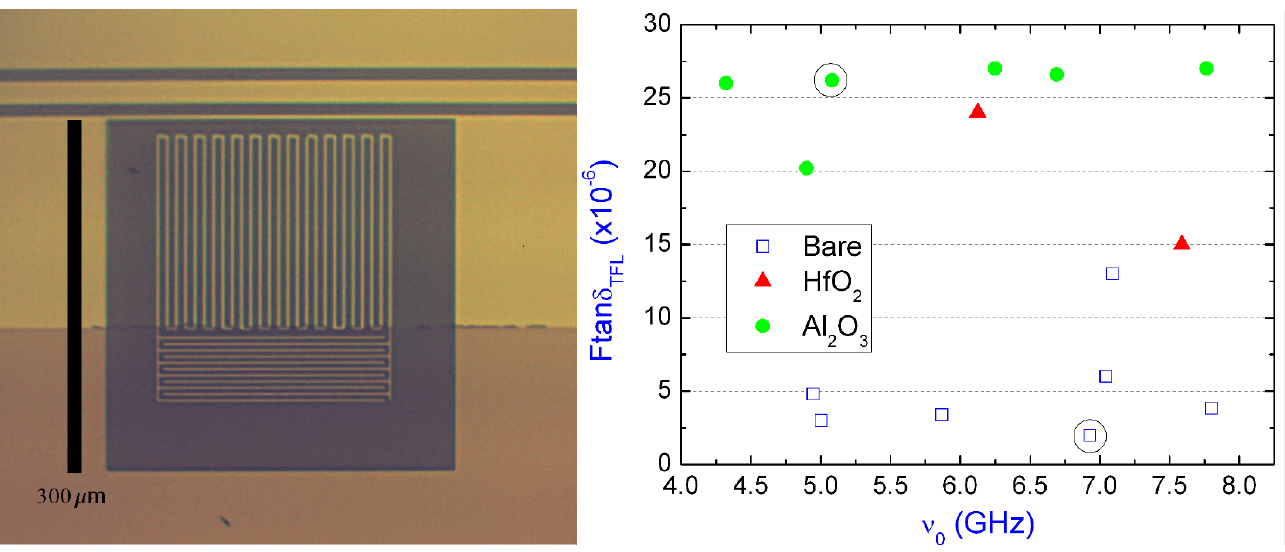}
\caption[short caption to go in LOF]{Left: Photograph of device, consisting of a long meandering inductive part and an inter-digitated capacitor. For covered resonators the capacitive part is coated by the additional dielectric layer (red strip). Right: Scatter plot comparing the loss tangents of bare resonators with those covered by $\text{Al}_{2}\text{O}_{3}$ and $\text{HfO}_{2}$. Plot highlights density of TLF's to be uniform with frequency and for covered resonators to have a loss tangent between 5-10 times higher than the bare resonator. Circled is a bare resonator, B1, at 6.93~GHz and an  $\text{Al}_{2}\text{O}_{3}$ covered resonator, A1, at 5.08~GHz which are featured in later measurements.}
\label{loss1}
\end{figure*}
Theoretical work\cite{Faoro2012} suggests that `slow' fluctuators can have a profound -but indirect- effect on the noise and losses in superconducting devices operating at microwave frequencies such as qubits. In simplified terms the model considers two distinct ensembles of TLFs: a coherent population that couples directly to the device, and a `slow' population which perturbs the coherent TLFs, by changing the tunnel splitting. Hence, whereas the coherent processes between the two energy levels are expected to have short time constants, they are in turn effectively being modulated by processes that can have time constants of the order of seconds or even hours.

 Motivated by QIP applications, recent studies on superconducting resonators have focused heavily on sources of dissipation. Measurements have evaluated the effects of magnetic fields\cite{Healey2008} and vortex motion\cite{Bothner2011}. The remaining dissipation channel is usually attributed to the presence of TLFs. The exact nature of TLFs remains contentious, but a variety of experiments have studied their effects\cite{Wang2009,Gao2008b,Vissers2012,lindstrom2009,macha2010} and recent models evaluated the contributions of TLFs in varying locations\cite{wenner2011}.



In this letter we study slow noise processes in low loss niobium (Nb) on sapphire resonators\cite{lindstrom2009}.
Resonators are by their very nature sensitive probes: any change in the environment will produce a change in the centre frequency $\nu_{0}$ of the resonator; making them ideal devices for studying noise. However, measuring this noise can be difficult due to the extrinsic low frequency noise present in equipment such as amplifiers and mixers\cite{Rubiolabook}. Here we overcome this problem by using a high-bandwidth measurement method based on a so-called Pound loop which operates at an offset frequency well above the extrinsic flicker corner frequency\cite{Lindstrom2011}. Additionally, by depositing a further dielectric layer on top of the resonator we study the effects of an increased TLF density.


Our samples consist of several lumped element resonators coupled to a common feed line (see fig. \ref{loss1}). Transmission through such resonators is described by $S_{21}={2}[{2+\frac{g}{1+2jQ_{l}x}}]^{-1}$ where $Q_{l}$ is the loaded quality factor (defined as the center frequency, $\nu_{0}$, divided by the bandwidth, $\Delta{\nu}$,  $Q_{l}=\nu_{0}/\Delta{\nu}$), $x$ is the fractional frequency shift $x=(\nu-\nu_{0})/\nu_{0}$ and g is the coupling parameter. The center frequency is defined by $\nu_{0}=(2\pi{\sqrt{(L+L_{K})C}})^{-1}$ where C is the capacitance, L the inductance and $L_{K}$ the kinetic inductance. $L_{K}$ varies with the penetration depth and hence with temperature and magnetic field\cite{
Healey2008}

Experimentally it is found that as temperature is reduced a monotic increase in center frequency is observed, while the quality factor increases due to reducing conductor losses as described by Mattis-Bardeen theory\cite{Mattis1958}. At temperatures much below the superconducting critical temperature, $T_{c}$, these mechanisms saturate. When lowering the temperature beyond this saturation a further change in both Q and center frequency is observed, which is well described in the theory of TLFs \cite{Schickfus1977} \cite{Strom1978}, by modeling a single TLF as a dipole that can shift states in an asymmetric well by thermally activated tunneling or absorption of resonant photons. The former effect leads to a change in the center frequency while the latter manifests as a power dependent Q. From TLF theory the change in permittivity can  be described by
\begin{equation}
\frac{\Delta\epsilon{(T)}}{\epsilon{(T_{0})}}=-\frac{2nd^{2}}{3\epsilon}
\left(
\text{ln}\left(\frac{T}{T_0{}}\right)-[g(T,\omega)-g(T_{0},\omega)]
\right)
\label{eps_vs_t}
\end{equation}
Where $\epsilon$ is the permittivity, \textit{n} the density of TLF states, \textit{d} the dipole moment, \textit{T} the temperature,  $T_{0}$ is a reference temperature and $g(T,\omega)=Re\Psi{(\frac{1}{2}+\hbar{\omega}/2\pi{i}k_{B}T)}$ and $\Psi$ is the complex digamma function. Changes in permittivity relate to frequency changes by $\frac{\Delta{\nu_{0}}}{\nu_{0}}=-\frac{F}{2}\frac{\Delta{\epsilon}}{\epsilon}$, with F being a geometric filling factor.

Equation \eqref{eps_vs_t} can be used to determine the \textit{intrinsic} loss tangent $F\text{tan}\delta{^{0}}_{TLF}=F\frac{2nd^{2}}{3\epsilon}$ of a resonator, which is proportional to the density of TLFs, \textit{n}, and their dipole moment $d^{2}$. This can differ slightly from the loss tangent determined by power dependent Q measurements, due to the former including the effect of non-resonant TLFs \cite{pappas2011}.
The presence of two distinct populations of TLF -`slow' and coherent- would remove the direct correlation between loss and noise in the resonator; the indirect effect of slow fluctuators means that they can influence the noise level of the resonator without affecting parameters such as the quality factor.


\begin{figure*}
\includegraphics{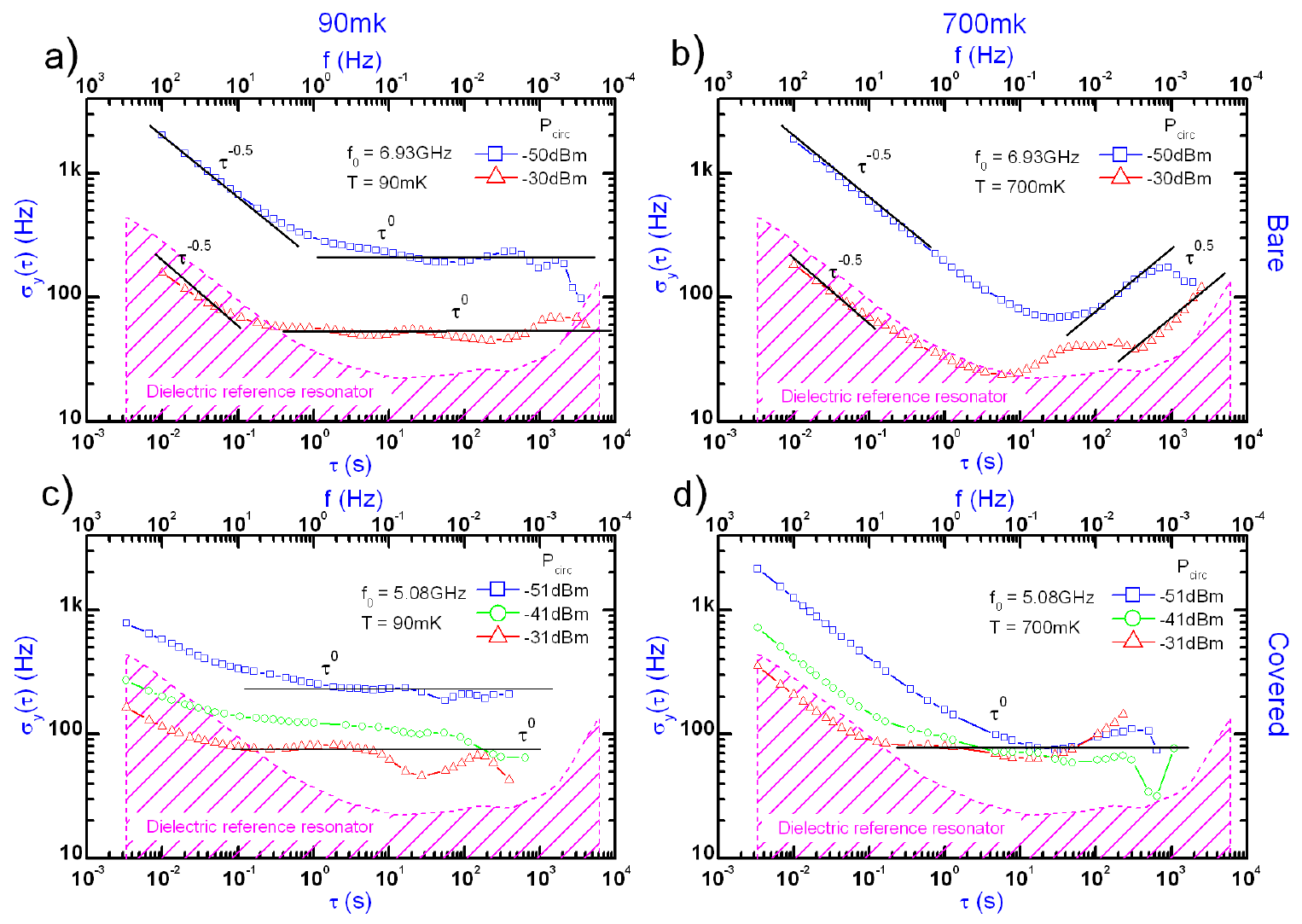}
\caption{RMS Frequency deviation as a function of measurement integration time. Successive shapes correspond to different microwave drive powers, triangles = -85~dBm, circles = -95~dBm and squares = -105~dBm. Plots a) and b) are of a bare resonator, B1, and an aluminium oxide covered resonator, A1, in plots c) and d) where the temperature is 90~mK for plots a) and c) and 700~mK for plots b) and d). }
\label{allan1}
\end{figure*}

A $\text{He}^{3}-\text{He}^{4}$ dilution refrigerator is used to measure samples, the input signal is attenuated by 50~dB and filtered by a 3.5~GHz high pass filter. The output signal is amplified by a cryogenic InP HEMPT amplifier with a noise temperature of 5~K and gain of 30~dB before further amplification at room temperature. The resonators consist of 200~nm thick niobium films deposited by RF sputtering on R-plane sapphire. To study the effects of TLFs some resonators are left bare, while others have a 50~nm layer of $\text{Al}_{2}\text{O}_{3}$ (or HfO$_2$) deposited by atomic layer deposition (ALD) over the inter-digitated capacitors (see Fig 1). The intrinsic loss tangent is measured by varying the temperature of the fridge between 50~mK and 800~mK. The frequency shift is tracked by a Pound loop (see ref \cite{Lindstrom2011}) and fit to equation \eqref{eps_vs_t} to extract the intrinsic loss tangent. Figure 1 shows the additional dielectric coating to consistently produce an increased loss tangent by a factor of 5-10 independent of resonance frequency. Circled is bare resonator, B1, and covered resonator, A1, which are studied further below.

The Pound loop uses phase modulation at 1~MHz which is above the 1/f corner frequency of the amplifier/mixer chain. Feedback is used to establish a frequency locked loop which tracks the center frequency of the resonator in real time within the loop bandwidth ($\sim$6~kHz). Readout of the feedback signal allows the centre frequency to be resolved to below the 10 Hz level, achieving fractional frequency resolution of 2 parts in $10^{9}$ for a 5~GHz resonator. Discriminating such small frequency shifts provides a vast improvement to both speed and accuracy of loss tangent measurements using eq. \eqref{eps_vs_t}. By directly tracking the centre frequency shift without need of fitting resonance data we make a significant reduction in measurement time. However, more interestingly, the high bandwidth of the feedback loop means the intrinsic frequency jitter of the resonator can also be directly measured by monitoring $\nu_{0}$ vs. time.

Analysis of the jitter can be performed in either the time or frequency domains. Frequency domain analysis produces spectral density plots ($S_{y}$ in units of $\text{Hz}^2/\text{Hz}$) however spectral analysis is not ideal for evaluating low frequency noise due to the numerical pole at 0 and windowing effects\cite{Rubiolabook} (See Supplemental Material at [URL] for [spectral analysis]). Instead time domain analysis is performed by the simple verifiable technique of calculating the Allan deviation ($\sigma_{y}(\tau)=\sqrt{\frac{1}{2}\langle{(\overline{y_{n+1}}-\overline{y_{n}})^{2}}\rangle{}}$ in units of $\text{Hz}$) which is similar to the standard deviation except that it converges for all noise processes. It hence serves as a measurement of the frequency jitter for a given measurement time. Both techniques describe noise processes obeying a power law, where for example flicker frequency noise is described by $S_{y}\propto{1/f^{1}}$ and $\sigma_{y}(\tau)\propto{\tau^{0}}$, see ref\cite{Lindstrom2011} for more information. A strong advantage of Allan analysis is to determine the time scales over which instrumental noise, e.g. from amplifiers dominates the noise, hence the Allan analysis determines the actual resolution obtainable within a given measurement bandwidth.


A dielectric resonator of diameter 14 mm and height 8 mm supported by a 14 mm post, with $Q_{l}\approx{10^{5}}$ and $\nu_{0}$=8.1~GHz is first measured at 90~mK as a reference to determine the extrinsic noise sources. Such resonators are known to be very stable and more importantly do not exhibit flicker frequency noise\cite{Hartnett2012}, instabilities instead arise due to thermal or mechanical fluctuations and appear as a linear frequency drift. The superconducting resonators are always at least as noisy as the dielectric resonator, in figure 2 the system noise floor is therefore labeled as the dielectric reference resonator and shown by the pink hashed region. The frequency jitter of a bare (covered) resonator, B1 (A1), with $Q_{l}\approx$35000, $\nu_{0}$=6.93~GHz and $F\text{tan}\delta^{0}_{TLF}$=2.0x$10^{-6}$ ($Q_{l}\approx$30000, $\nu_{0}$=5.08~GHz and  $F\text{tan}\delta^{0}_{TLF}$=2.6x$10^{-5}$) is shown in figure 2a (figure 2c).

Under high applied drive the circulating power exceeds the saturation power\cite{macha2010}, which is the typical regime for kinetic inductance detector applications \cite{leduc2010}. These high powers should saturate many resonant TLFs leaving the frequency shift to be caused by non-resonant TLFs \cite{pappas2011}.

At low temperatures and low microwave drive the behaviour of both the bare and covered resonators is similar (square traces in figures 2a and 2c). There are two dominant regions in these plots, at short times a large jitter is present due to a frequency independent noise process which manifests as white frequency noise (noise obeying a power law of $\tau^{-0.5}$) which is predominately caused by the instrumentation \cite{whitenoisecomment}. The second region occurs at long time scales from 1 second. Here we directly observe flicker frequency noise (noise obeying a power law of $\tau^{0}$) which limits frequency stability in bare resonators to $\approx$200~Hz and in covered resonators to $\approx$300~Hz.

When increasing the microwave drive the entire noise level is observed to decrease for both the bare and covered resonators (triangle traces in figures 2a and 2c). The noise processes remain the same, white frequency noise at short times and flicker frequency noise at long time scales. Increased microwave drive saturates the resonant TLFs leading to a reduction in the flicker frequency noise level by a factor of $\approx$3. At 90~mK the thermal energy is not sufficient to excite TLFs, in figures 2b and 2d measurements are performed at 700~mK where TLFs can be thermally excited.

At higher temperatures and low microwave drive the behaviour of covered resonators is similar to that at low temperatures, with white frequency noise dominating at short times and flicker frequency noise dominating at times in excess of 10 seconds. Compared to the bare resonator, white frequency noise still dominates at short times but the longer time scales now exhibit random frequency walk noise (noise obeys a power law of $\tau^{0.5}$). The lack of flicker frequency noise leads to a reduced noise level with minimum of 90~Hz when averaging for 30 seconds. At higher microwave drive the flicker frequency noise level of the covered resonator remains the same, while the bare resonator sees a consistently lower noise level, again without flicker frequency noise. The lowest noise level is found to be 30~Hz, corresponding to a fractional frequency shift of 4 parts in $10^{9}$ when averaging for 5 seconds.

These resonators are shown to exhibit flicker frequency noise, consistent with the resonator coupling to a bath of two level fluctuators, if the TLF ensemble is coherent and weakly interacting with the resonator such a bath has been theoretically shown to produce a flicker frequency spectral noise\cite{Shnirman2005}. The noise level is also found to increase with increasing loss tangent, consistent with noise increasing with increasing density of TLFs\cite{Shnirman2005}\cite{barends2008}. At low temperatures there exists a power dependence of the noise which could scale similar to the $P_{int}^{-1/2}$ dependence suggested by Gao et al\cite{Gao2008c} however more measurements are required to verify the exact power dependence.

At higher temperatures the flicker frequency is not observed for bare resonators, considering the loss tangent data, we expect 700~mK to exceed the $\Delta_{max}$ energy scale of the TLFs determined by Shnirman et al\cite{Shnirman2005}, resulting in the resonator no longer coupling to the coherent TLF ensemble and hence not exhibiting flicker frequency noise. A possible candidate for the random walk noise process is thermally activated flux motion. Within the covered resonator flicker frequency noise exists even at 700~mK, however the level is no longer power dependent, in contrast to measurements of `fast' noise processes\cite{kumar2008}. It is possible that the surface TLFs have a larger $\Delta_{max}$ than those in the niobium-sapphire interface, but this should appear within the loss tangent measurements. Instead we suspect the resonator couples to a coherent bath of TLFs, which are themselves affected by the incoherent action of other TLFs as was recently suggested by Faoro and Ioffe\cite{Faoro2012} and observed by Grabovskij\cite{Grabovskij2012}. The action of `slow' fluctuators limits the RMS deviation of the resonance frequency over large time scales $\sim$1 second which leads to the observed flicker frequency noise. Since these `slow' fluctuators are not coherently coupled to the resonator, they are not affected by microwave drive. 

We contrast our work with previous experiments on superconducting resonators which used a homodyne technique\cite{Gao2007a}\cite{kumar2008}\cite{barends2008}, where a short measurement time makes it difficult to accurately distinguish between effects from instrumentation and any intrinsic slow noise processes. We therefore emphasize the need to both measure the \emph{instrumental} contribution and to measure for long time scales to retain statistical confidence when analyzing effects at low frequencies. Previous measurements identified a noise spectrum scaling as $S_{y}\propto f^{-0.5}$, as opposed to the unequivocal observation of the flicker noise $S_{y}\propto{f^{-1}}$ in our work [see supplementary material]. 

The $S_{y}\propto f^{-0.5}$ noise was also seen in our early development\cite{Lindstrom2011}, however improvements in our experimental setup led to the process not being observed in this work\cite{mixercomment}. We stress that the standard power law noise model assumes a noise spectrum scaling as $S_{y}\propto1/f^{\alpha}$ where $\alpha$ is an integer value\cite{Rubiolabook} and each value of $\alpha$ corresponds to a named and identifiable process, eg. $\alpha=1$ observed in this work corresponds to a flicker FM process. In this regard this work importantly brings noise measurements on superconducting resonators in line with those on other superconducting devices\cite{Koch1980} which also fit within the standard power law noise model. 

We note that measurements at intermediate frequencies below the resonator cut off (Leeson) frequency (Leeson frequency $f_{L}=\nu_{0}/2Q_{l}$), results in a phase-to-frequency conversion of noise\cite{Rubiolabook} and hence previous homodyne noise measurements were often mis-interpreted as the \emph{phase} noise spectra, $S_{\phi}$, when infact they showed the \emph{frequency} noise spectra $S_{y}\propto{f^{-0.5}}$.

To conclude we directly observed flicker frequency noise - a  slow process in superconducting resonators and highlight different behavior in power, temperature and TLF density. In particular the flicker level is observed to decrease with increasing temperature and increasing microwave drive. These slow processes are described in theory\cite{Shnirman2005}\cite{Faoro2012} and have been shown to affect qubits\cite{Grabovskij2012} and are expected to affect other superconducting or charge sensitive devices such as kinetic inductance detectors.


The authors acknowledge V. Antonov, C. Shelly, G. Ithier, G. Tancredi, R. Fraizer and J. Gallop for fruitful discussions. This work was supported by the NMS Pathfinder program and the EPSRC.








\bibliographystyle{nature}      			
\bibliography{newbib}

\appendix


\end{document}